\documentclass[12pt]{article}
\usepackage{graphicx}
\newcommand {\bi} {\bibitem}
\newcommand {\be} {\begin{equation}}
\newcommand {\ee} {\end{equation}}
\newcommand {\bea} {\begin{eqnarray} }
\newcommand {\eea} {\nonumber \end{eqnarray}}
\newcommand {\eps} {\epsilon}
 \newcommand {\si} {\sigma}

\newcommand {\ga} {\gamma}

 \newcommand {\al} {\alpha}

\newcommand {\ba} {\overline}
\newcommand {\lan} {\langle}
\newcommand {\ran} {\rangle}

\newcommand {\cP}  {{\cal P}}

\newcommand {\bc} {\begin{center}}
\newcommand {\ec} {\end{center}}
\newcommand {\bd}{\begin{displaymath}}
\newcommand {\ed}{\end{displaymath}}

\newcommand {\for} {\ \ \ \mbox{for}\ \ }

\def \form#1 {eq. (\ref{#1}) }
\def \parziale#1#2  {{\partial {#1} \over \partial {#2}}}

\begin{document}

\title{The physical Meaning of Replica Symmetry Breaking}
\author{Giorgio Parisi \\ Dipartimento di Fisica, Sezione INFN and unit\`a INFM,\\
Universit\`a di Roma ``La Sapienza'',\\
Piazzale Aldo Moro 2,
I-00185 Roma (Italy)\\
giorgio.parisi@roma1.infn.it}
\maketitle
\abstract{In this  talk I will presente the physical meaning of replica symmetry breaking
stressing the physical concepts. After introducing the theoretical framework and the experimental 
evidence for replica symmetry breaking, I will describe some of the basic ideas  using a 
probabilistic language. The predictions for off-equilibrium dynamics will be shortly outlined.  }
\section{Introduction}

In this talk I will underline the physical meaning of replica symmetry breaking 
\cite{MPV,PABOOK,CINQUE}. I will
stress the physical concepts and I will skip most of the technical details. It is an hard 
job because the field has grown in a significant way in the last twenty years and many 
results are available.

I will try to make a selection of the most significant results, which is however  partly arbitrary 
and incomplete. The main points I would like to  discuss are:

\begin{itemize}
    
\item Complex Systems: the coexistence of many phases.
\item The definition of the overlap and its probabilities distributions.
\item Experimental evidence of replica symmetry breaking.
\item High level statistical mechanics.
\item Stochastic stability.
\item Overlap equivalence and ultrametricity.
\item The algebraic replica approach.
\item Off-equilibrium dynamics.

\end{itemize}

As you can see from the previous list, in this talk I will try to connect rather different topics 
which can be studied using an unified approach in the replica framework.

\section{Complex Systems: the coexistence of many phases}

Boltzmann statistical mechanics can be considered an example of a successful {\sl redutionistic} 
program in the sense that it gives an microscopic derivation of the presence of emergent 
(collective) behaviour  of a system which has  many variables.  This phenomenon is known as phase 
transition.

If the different phases are separated by a first order transition, just at the phase transition 
point  a very interesting phenomenon is present: phase coexistence.  This usually happens if we 
tune one parameter: the gas liquid coexistence is present on a line in the pressure-volume plane, 
while the liquid-gas-solid triple point is just a point in this plane.  This behaviour is summarized 
by the Gibbs rule which states that, in absence of symmetries, we have to tune $n$ parameters in 
order to have the coexistence of $n+1$ phases.

The Gibbs rule is appropriate for many systems, however in the case of complex systems we 
have that the opposite situation is valid: {\sl the number of phases is very large 
(infinite) for a generic choice of parameters}.  This last property may be taken as a 
definition of a complex system.  It is usual to assume that all these states are globally 
very similar: translational invariant quantities (e.g. energy) have the same value in all 
the phases (apart from corrections proportional to $ N^{-1/2}$), this last properties 
being called phase democracy.  These states cannot be separated by external parameters 
coupled to translational invariant quantities, but only by comparing one state with an other.

An example of this phenomenon would be a very long heteropolymer, e.g. a protein or  RNA,
which may folds in many different structures. However   quite different foldings may have 
a very similar density. Of course you will discover that two proteins have folded in two 
different structures if we compare them.

\begin{figure}
    \centering
\includegraphics[width=0.6\textwidth]{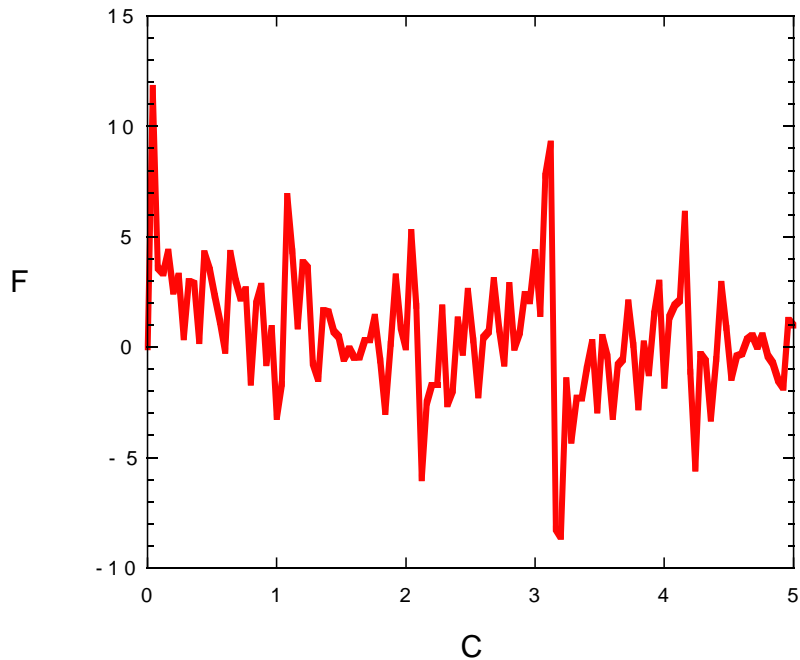}
\caption{An artistic view of the free energy of a complex system as function of the 
configuration space.}
\label{ART}
\end{figure}

In order to be precise we should consider a large but ({\sl finite}) system.  We want to 
decompose the phase state in valleys (phases, states) separated by barriers \footnote{For 
a discussion of the meaning of finite volume states, which are different from infinite 
volume states \cite{NS}, see \cite{CINQUE}.}.
If the free 
energy as function of the configuration space has many minima (a corrugated free energy 
landscape, as shown in fig. (\ref{ART})) the number of states will be very large. An analytic and 
quantitative study  of the properties of the free energy landscape in a particular model can be 
found in \cite{CGP}.

Let us consider for definitiveness  a spin system with $N$ points
(spins are labeled by $i$, which in some cases will be a lattice point).

States (labeled by $\alpha$) are characterized by different local  magnetizations:
\be  
m_{\alpha}(i) =\lan \sigma(i) \ran_{\alpha}  \  ,
\ee
where $\lan \cdot \ran_{\alpha}$ is the expectation value
in the valley labeled by $\alpha$.
The average done with the Boltzmann distribution is denoted as $\lan \cdot \ran$ and it 
can be written as linear combinations of the averages inside the valleys. 
We have the relation:
\be
\lan \cdot \ran \approx \sum_{\alpha} w_{\alpha} \lan \cdot \ran_{\alpha}\ . \label{DECON}
\ee
We can write that the relation
\be
w_{\alpha} \propto  \exp(-\beta F_{\alpha}) \ ,
\ee
where by definition $F_{\alpha}$ is the free energy of the valley labeled by $\alpha$.

In the rest of this talk we will call $J$ the control parameters of the systems. The 
average over $J$  will be denoted by  a bar (e.g. $\ba{F}$).
In the cases I will consider here a quenched disorder is present: 
the variables $J$ parametrize  the quenched disorder.

The general problem we face is to find those quantities which {\sl do not} depend on $J$ and to find 
the probability distribution of those quantities which {\sl do} depend on $J$.

\section{The overlap and its probabilities}

As we have already remarked in the case of heteropolymers folding, states may be 
separated making a comparison among them. At this end it is convenient to consider their 
mutual overlap. 
Given two configurations ($\sigma$ and $\tau$), we define their overlap:
\be
q[\sigma,\tau]= \frac{1}{N} \sum_{i=1,N}\sigma(i)\tau(i) \ .
\ee
The overlap among the states is defined as
\be
q(\alpha,\gamma)=\frac{1}{N} \sum_{i=1,N}m_{\alpha}(i)m_{\gamma}(i)\approx
q[\sigma,\tau]\ ,
\ee
where $\sigma$ and $\tau$ are two generic configurations that belong to the states $\alpha$ 
and $\gamma$ respectively.

We define $P_{J}(q)$ as probability distribution of the overlap  $q$ at given $J$, 
i.e. the histogram of $q[\si,\tau]$ where $\si$ and $\tau$ are two equilibrium 
configurations. Using eq. (\ref{DECON}), one finds that

\be P_{J}(q)=\sum_{\alpha,\gamma}w_{\alpha}w_{\gamma}\delta(q-q_{\alpha,\gamma}) \ ,
\ee
where in a finite volume system the delta functions are smoothed.
If there is more than one state, $P_{J}(q)$ is not a single delta function
\be
P_{J}(q)\ne \delta (q-q_{EA}) \ .
\ee
If this happens we say that the replica symmetry is broken: two identical replicas of the 
same system may stay in a quite different state.

There are many models where the function $P_{J}(q)$ is non-trivial: 
a well known example is given by Ising spin glasses \cite{MPV,BY,FiHe}. In this case the Hamiltonian
is given by

\be H=-\sum_{i,k}J_{i,k}\sigma_{i}\sigma_{k} -\sum_{i}h_{i} \si_{i}\ ,
\ee
where $\si=\pm 1$ are the spins. The variables $J$ are random couplings (e.g. Gaussian or 
$\pm 1$) and the variables $h_{i}$ are the magnetic fields, which may be point dependent.

Let is consider two different models for spin glasses:
\begin{itemize}
\item The Sherrington Kirkpatrick model (infinite range): all $N$ points are connected: 
$J_{i,k}=O(N^{-1/2})$.  Eventually $N$ goes to infinity.

\item  Short range models: $i$ belongs to a $L^{D}$ lattice. The interaction is nearest neighbour 
(the variables $J$ are or zero or of order 1) and eventually 
 $L$  goes to infinity at fixed $D$ (e.g. $D=3$).

\end{itemize}

Analytic studies have been done in the case of the SK model, where one can prove 
rigorously that the function $P_{J}(q)$ is non-trivial. In the finite dimensional case no
theorem has been proved and in order to answer to the question if the function $P_{J}(q)$
is trivial we must resort to  numerical simulations \cite{romani-young} or to 
experiments.

\begin{figure}
    \centering
  \includegraphics[width=0.33\textwidth,angle=270]{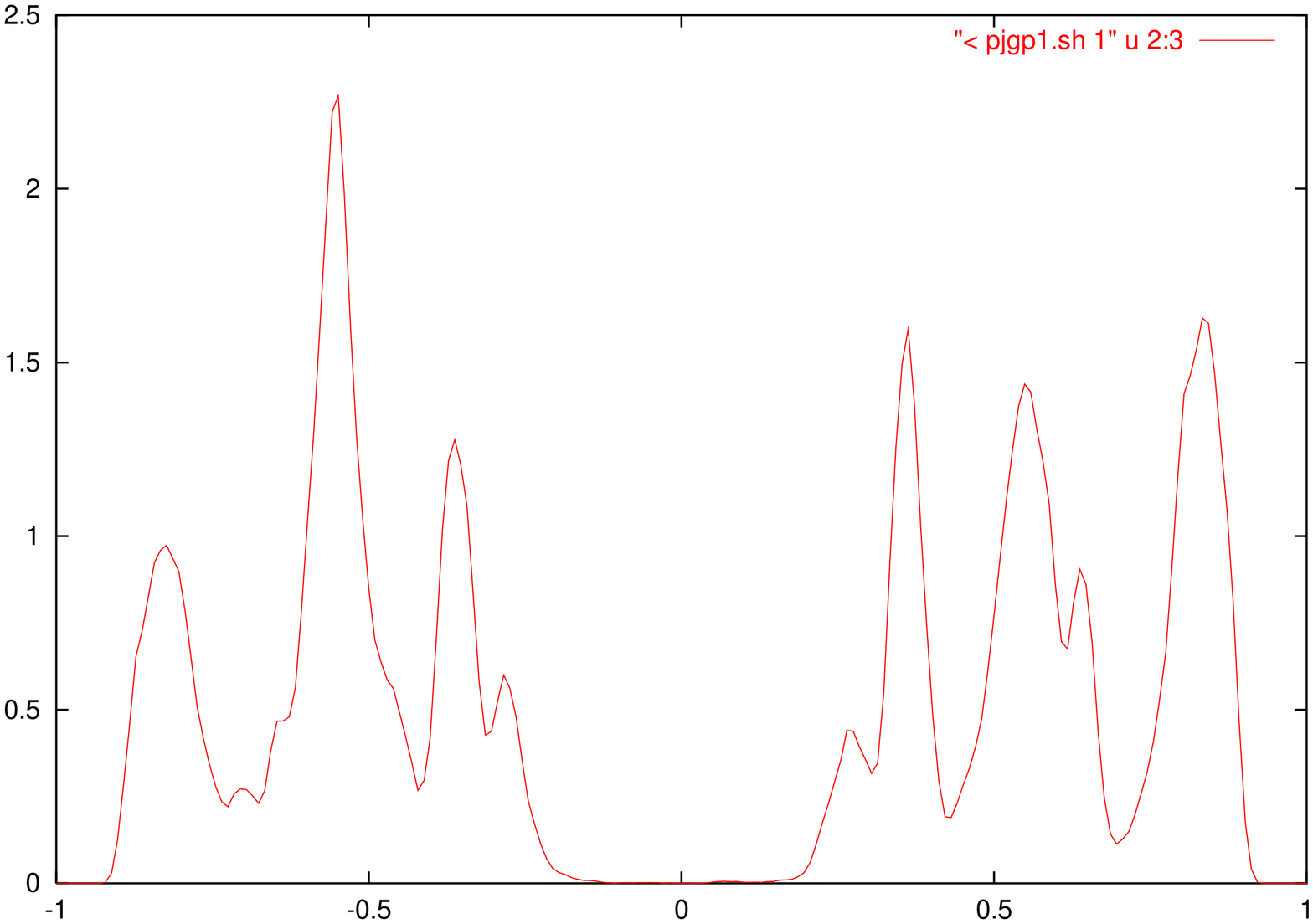}
  \includegraphics[width=0.33\textwidth,angle=270]{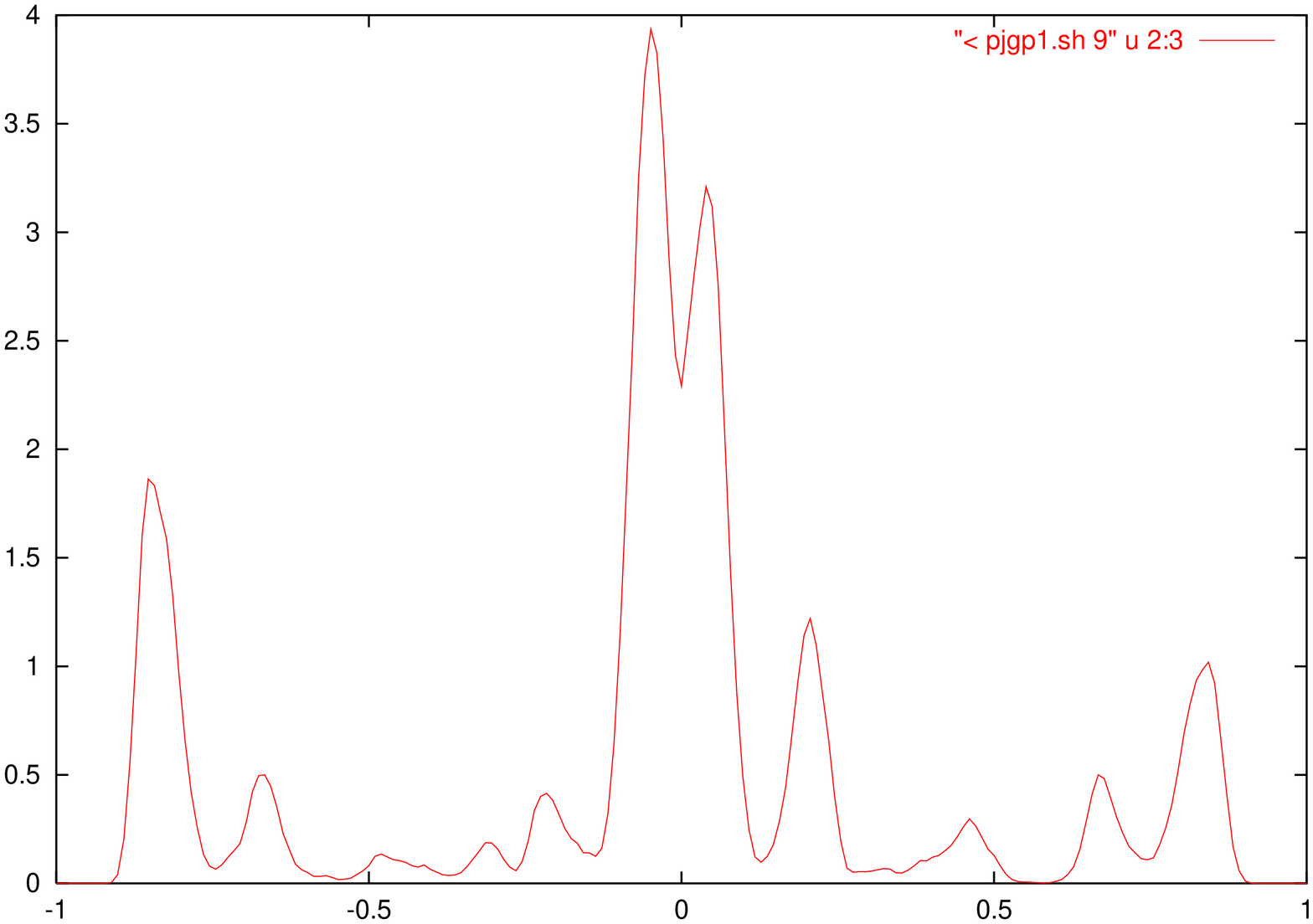}

  \includegraphics[width=0.33\textwidth,angle=270]{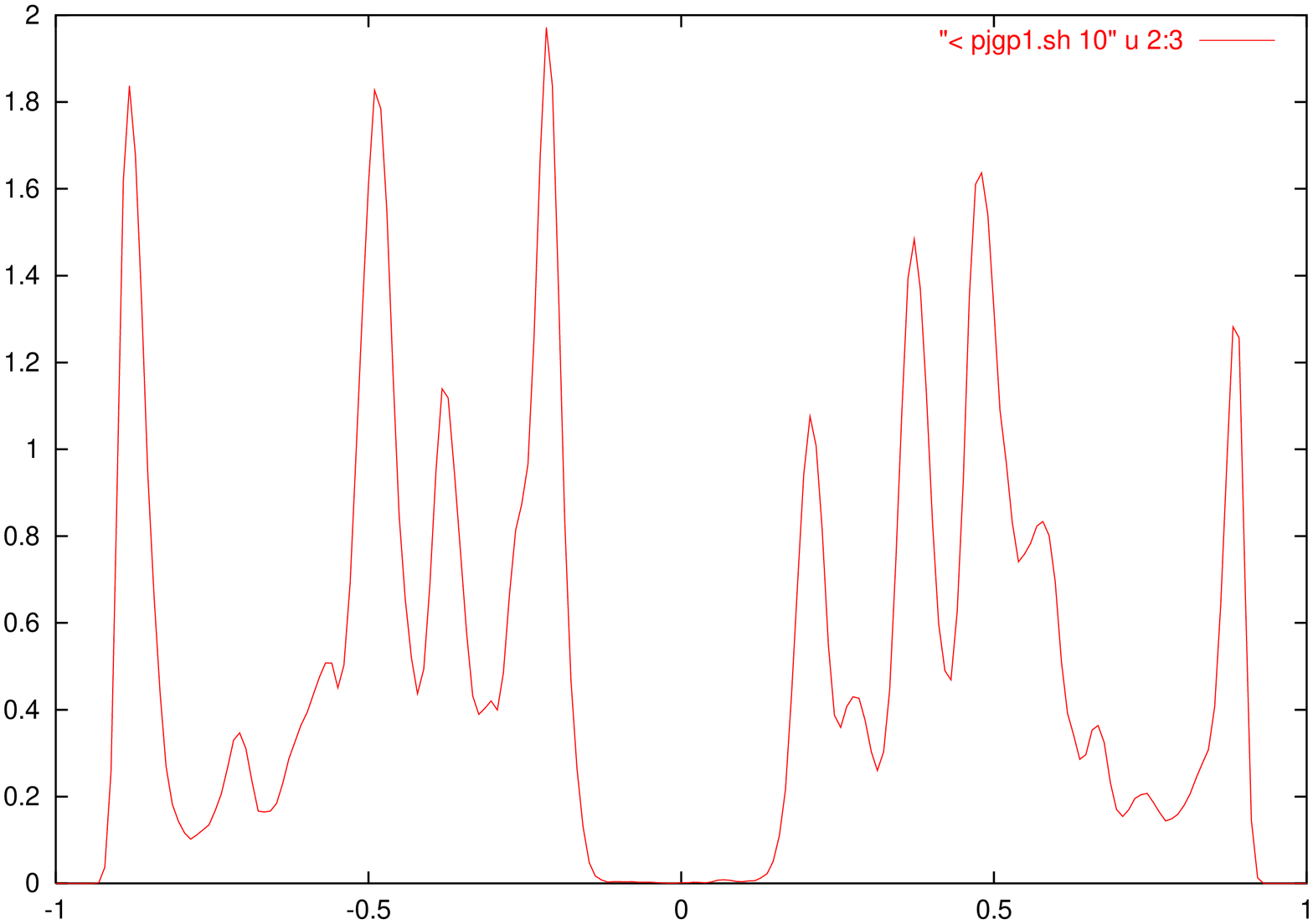}
  \includegraphics[width=0.33\textwidth,angle=270]{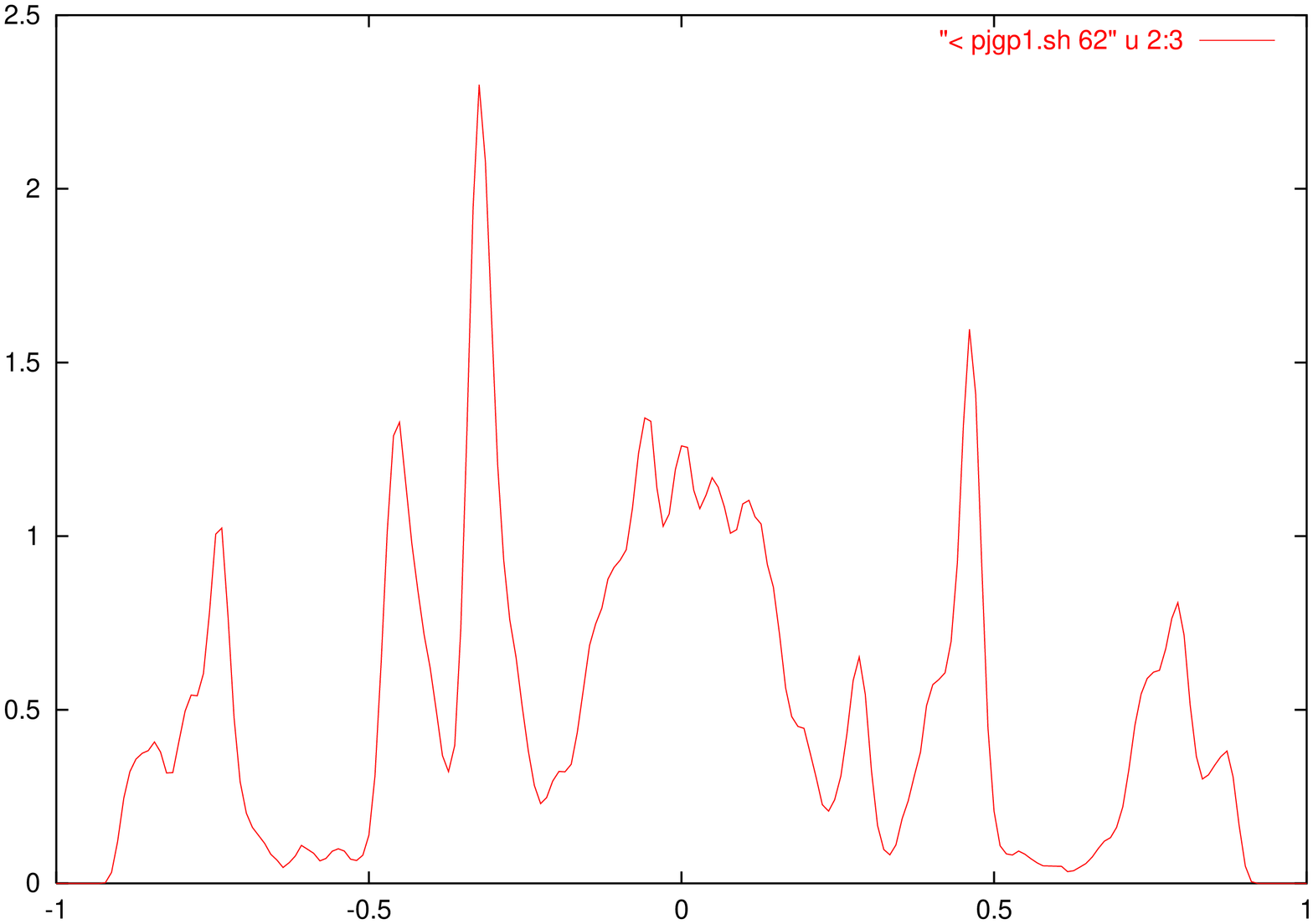}
\caption{The function $P_{J}(q)$ for four different 
samples (i.e different choices of $J$) for $D=3\ \ \ L=16$ ($16^{3}$ spins).}
\label{FEW}
\end{figure}
In fig. (\ref{FEW}) we show the numerical simulations for 4 different systems (i.e. 
different choices of the $J$ extracted with the same probability) of size 
$16^{3}$ \cite{XXX}. The slightly asymmetry of the functions is an effect of the finite 
simulation time. It is evident that the function $P_{J}(q)$ is non-trivial and it looks 
like a sum a smoothed delta functions. It is also evident that the function $P_{J}(q)$ 
changes dramatically from system to systems.

It is interesting to see what happens if we average over the samples. We can this define
\be
P(q)=\ba{P_{J}(q)} \ .
\ee
Of course,  if $P_{J}(q)$ depends on $J$, we have that 
\be
\ba{P_{J}(q_{1})P_{J}(q_{2})}\equiv P(q_{1},q_{2})\ne P(q_{1})P(q_{2}) \ .
\ee

\begin{figure}
    \centering
  \includegraphics[width=.7\textwidth]{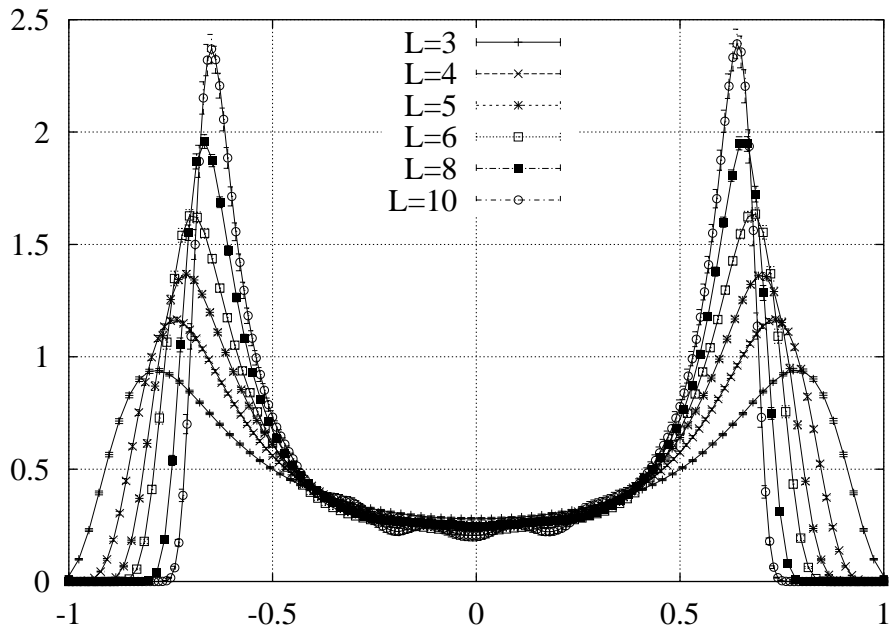}
\caption{The function $P(q)=\ba{P_{J}(q)}$ after average over many samples (D=4, L=3\ldots 10) .}
\label{D4}
\end{figure}

In fig. (\ref{D4}) we show the average over many samples of $P_{J}(q)$ in the $D=4$ case (a similar 
picture holds in $D=3$).  In this way we obtain a smooth function, with two picks which 
are slightly shifted and becomes sharper and sharper when the size of the system becomes 
larger.  It seems quite reasonable that when the system becomes infinite this peak evolves toward a 
delta function which corresponds to the contribution coming  from  two configurations $\sigma$ and $\tau$ 
which belongs to the same state.
\section{Experimental evidence of replica symmetry breaking}
Replica symmetry breaking affects the equilibrium properties of the 
system and in particular the magnetic susceptibility. For example let us consider a 
system  in presence of an external constant magnetic field, with Hamiltonian given by:
\be
H[\si]=H_{0}[\si]+\sum_{i}h\si_{i}\ .
\ee
\begin{figure}
    \centering
  \includegraphics[width=.6\textwidth]{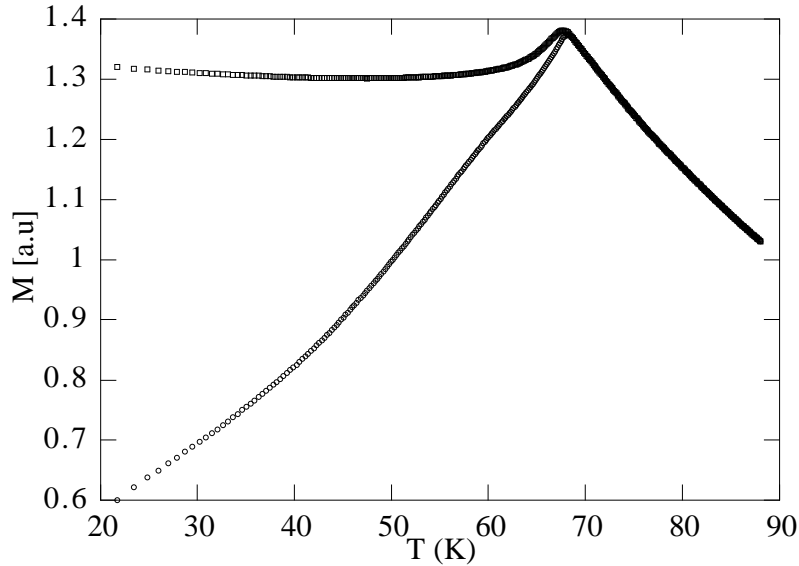}
\caption{ FC- and ZFC-magnetisation (higher and lower curve respectively) vs.  temperature of Cu(Mn13.5\%), $H$ =1 Oe (taken from 
\cite{EXP}).  For a such a low field the magnetization is proportional to the susceptibility.}
\label{2CHI}
\end{figure}

As soon as replica symmetry is broken we can define two magnetic susceptibilities which 
are different:
\begin{itemize}
    \item The magnetic susceptibility that we obtain when the system is constrained to remain in a 
    valley.  In the limit of zero magnetic field this susceptibility is given by 
    $\chi_{LR}=\beta(1-q_{EA})$.
    
    \item The total susceptibility magnetic susceptibility (the system 
    is allowed to change state as an effect of the magnetic field).  In the limit of zero magnetic 
    field this susceptibility is given by $\chi_{eq}=\beta\int dq \ P(q) (1-q) \le \beta(1-q_{EA})$ .
\end{itemize}
Both susceptibilities are experimentally observable. 
\begin{itemize}
    \item The first susceptibly is the susceptibly that you measure 
if you add an very small magnetic field at low temperature.  The field should be small enough in 
order to neglect non-linear effects.  In this situation, when we change the magnetic field, the 
system remains inside a given state and it is not forced to jump from a state to an other state and 
we measure the ZFC (zero field cooled) susceptibility, that corresponds to $\chi_{LR}$.  
 \item
The second susceptibility can be 
approximately measured doing a cooling in presence of a small field: in this case the system has the 
ability to chose the state which is most appropriate in presence of the applied field.  This 
susceptibility, the so called FC (field cooled) susceptibility is nearly independent from the 
temperature and corresponds to $\chi_{eq}$.
\end{itemize}

Therefore one can identify $\chi_{LR}$ and $\chi_{eq}$ with the ZFC susceptibility and with the
FC susceptibility respectively. The experimental plot of
the two susceptibilities is shown in fig. (\ref{2CHI}). They are clearly equal in the high 
temperature phase while they differ in the low temperature phase. 

The difference among the two susceptibilities is a crucial signature of replica symmetry breaking 
and, as far as I known, can explained only in this framework.  This phenomenon is due to the fact 
that a small change in the magnetic field pushes the system in a slightly metastable state, which 
may decay only with a very long time scale.  This may happens only if there are many  states 
which differs one from the other by a very small amount in free energy.

\section{The theoretical framework}

The general theoretical problem we face is to find out which is the probability distribution of the 
set of all $q_{\alpha,\gamma}$ and $F_{\alpha}$ (or equivalently $w_{\alpha}$).  More precisely for 
each given $N$ and $J$ we call ${\bf P}$ the set of all $q_{\alpha,\gamma}$ and $F_{\alpha}$: as we 
have seen this quantity has strong variations when we change the system.  We now face the problem of 
compute the probability distribution of ${\bf P}$, that we call $\cP({\bf P})$.  Moreover it should 
be clear that also when $\cP({\bf P})$ is known the computation of the average of some quantities 
over this distribution is non-trivial because for large systems ${\bf P}$ contains an unbounded 
number of variables.  The task of doing these kind of averages we can regarded as a sort of high 
level (macroscopic) statistical mechanics \cite{FACE}, where the basic entities are the phases of 
the system, while the usual statistical mechanics can be thought as low level (microscopic) 
statistical mechanics \footnote{The words ``low level'' and ``high level'' are used in the same 
spirit as ``low level language'' and ``high level language'' in computer science.}.

The number of possible form of the probability distribution $\cP({\bf P})$ is very high 
($\bf P$ is an infinite dimensional vector). In order to reduce the number of possible 
distributions one usually uses some general guiding principles.    
    The simplest theory is 
based on two principles: 
\begin{itemize}
    \item Stochastic stability \cite{guerra,aizenman,PARI}.
    \item Overlap equivalence \cite{PARI,ULTRA}. 
\end{itemize}

Stochastic stability is nearly automatically implemented in the algebraic replica approach 
that will be described in the next section and it seems to be a rather compulsory property in 
equilibrium statistical mechanics.
Overlap equivalence is usually implemented in the algebraic replica approach, but is 
certainly less compulsory than stochastic stability.

\subsection{Stochastic stability}
In the nutshell stochastic stability states that the system  we are considering behaves
like a generic random system.
Technically speaking in order to formulate stochastic stability we have to consider the 
statistical properties 
of the system with Hamiltonian given by the original Hamiltonian ($H$) plus a random 
perturbation ($H_{R}$):
\be
H(\eps)=H+\eps H_{R} \  .
\ee
Stochastic stability states that all the properties of the system are smooth functions
of $\eps$ around $\eps=0$, after doing the appropriate
averages
over the original  Hamiltonian and the random Hamiltonian.

Typical examples of random perturbations perturbations (we can chose the value of  $r$ in 
an arbitrary way):

\begin{equation}
H_{R}^{(r)}=N^{(r-1)/2}\sum_{i_{1}\ldots i_{r}} R({i_{1}\ldots i_{r}}) \si(i_{1}) \ldots 
\si(i_{r}) \ ,
\end{equation}
where for simplicity we can restrict ourselves to the case where the variables $R$ are 
random uncorrelated Gaussian variables.
For $r=1$ this perturbation corresponds to adding a random magnetic field:
\begin{equation}
H_{R}^{(1)}=\sum_{i_{1}} R({i_{1}}) \si(i_{1}) \ .
\end{equation}

Stochastic stability is non-trivial statement:  when we add a perturbation the weight of 
the states changes of an amount that diverges when $N$ goes to infinity at fixed $\eps$.
Indeed the variation in the individual free energies is given by
$
\delta F_{\alpha}=O(\eps N^{1/2}) \ .
$

It useful to remark that if a symmetry is present, a system cannot be stochastically stable.  Indeed 
spin glasses may be stochastically stable only in the presence of a finite, non-zero magnetic field 
which breaks the $\si \leftrightarrow -\si$ symmetry.  
If a symmetry is present, stochastic stability may be valid only for those quantities which are invariant 
under the action of the symmetry group.  It is also remarkable that the union of two non-trivial 
uncoupled stochastically stable systems is {\sl not} stochastically stable.  Therefore a non-trivial 
stochastically stable system cannot be decomposed as the union of two or more parts whose 
interaction can be neglected.

We will now give an example of the predictive power  of stochastic stability \cite{QUATTRO}.

There are systems in which the replica symmetry is broken at one-step. In other words in 
thus kind of systems the 
overlap may take only two values:
\be
q_{\alpha,\alpha}=q_{1}=q_{EA}, \ \ \ q_{\alpha,\gamma}=q_{0} \for \alpha\ne \gamma  \ .
\ee
This is the simplest situation: all pairs of different states have the same (i.e. $q_{0}$) 
mutual overlap, which for simplicity we will take equal to zero.
The only quantity we have to determine is the probability of the free energies. The free energies are 
assumed to be random uncorrelated variables and the 
the probability of having a state with total free energy in the interval $[F, F+dF]$ is 
\be
\rho(F)dF \ .
\ee
Stochastic stability implies that in the region which is dominant for the thermodynamic 
quantities, i.e. for the states having not too high free energy:
\be
\rho(F)\propto \exp(\beta m (F-F_{0}))\ ,\label{ONESTEP}
\ee
where $F_{0}$ a system dependent reference free energy, proportional to $N$.
As a byproduct the function $P(q)$ can be magnetisations computed and is given by
\be
P(q)=m \delta(q)+(1-m) \delta(q-q_{EA})\ .
\ee

\begin{figure}
    \centering
\includegraphics[width=0.65\textwidth]{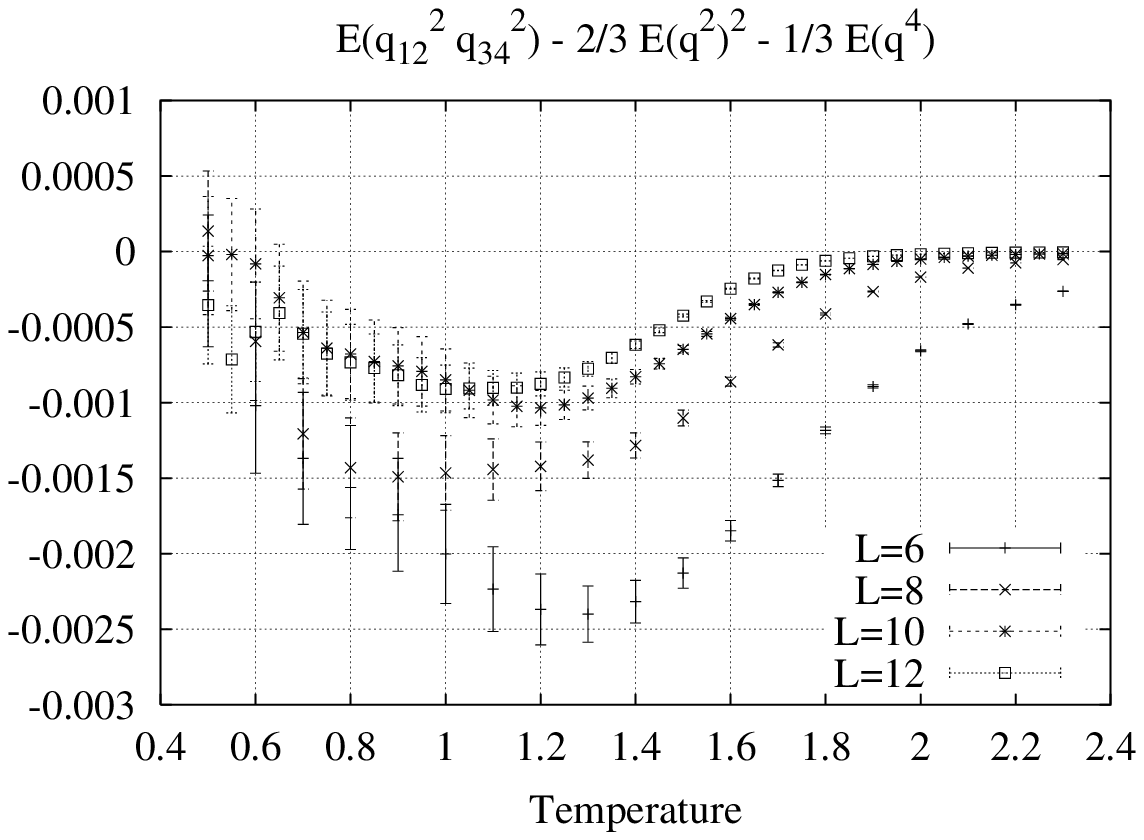}
\caption{The quantity $\ba{\lan q^2 \ran^{2}} - (\frac13 \ba{\lan q^4\ran}+
\frac23 \ba{\lan q^2\ran}^2)$ as function of the temperature for different values of $L$ in $D=3$.} 
\label{G1}
\end{figure}

The proof of the previous statement is rather simple \cite{QUATTRO}.  Stochastic stability imposes 
that the form of the function $\rho(F)$ remains unchanged (apart from a possible shift in $F_{0}$) 
when one adds a small random perturbation.  Let us consider the effect of a perturbation of strength 
$\epsilon$ on the free energy of a state, say $\alpha$.  The unperturbed value of the free energy is 
denoted by $F_\alpha$.  The new value of the free energy $G_{\al}$ is given by
\be
G_{\al}=F_{\al}+\eps r_{\al}\ ,
\ee
where $ r_{\al}$ are identically distributed
uncorrelated random numbers.
Stochastic stability implies that the distribution
$\rho(G)$ is the same as $\rho(F)$. \
Expanding to
second order in $\epsilon$ we get:
\be
\frac{d\rho}{dF}\propto \frac{d^2\rho}{dF^2}\ .
\ee
Therefore $\rho(F)\propto \exp(\beta m (F-F_{0}))$.

In the general case stochastic stability implies  that
\be
P(q_{1},q_{2})\equiv\ba{P_{J}( q_{1}) P_{J}(q_{2})} 
 = \frac23 P(q_{1}) P(q_{2}) +
\frac13 P(q_{1}) \delta(q_{1}-q_{2}) \ . 
\ee
A particular case of the previous relation is the following one
\be
\ba{\lan q^2 \ran^{2}}= \frac13 \ba{\lan q^4\ran}+
\frac23 \ba{\lan q^2\ran}^2\ .
\ee

\begin{figure}
    \centering
\includegraphics[width=0.65\textwidth]{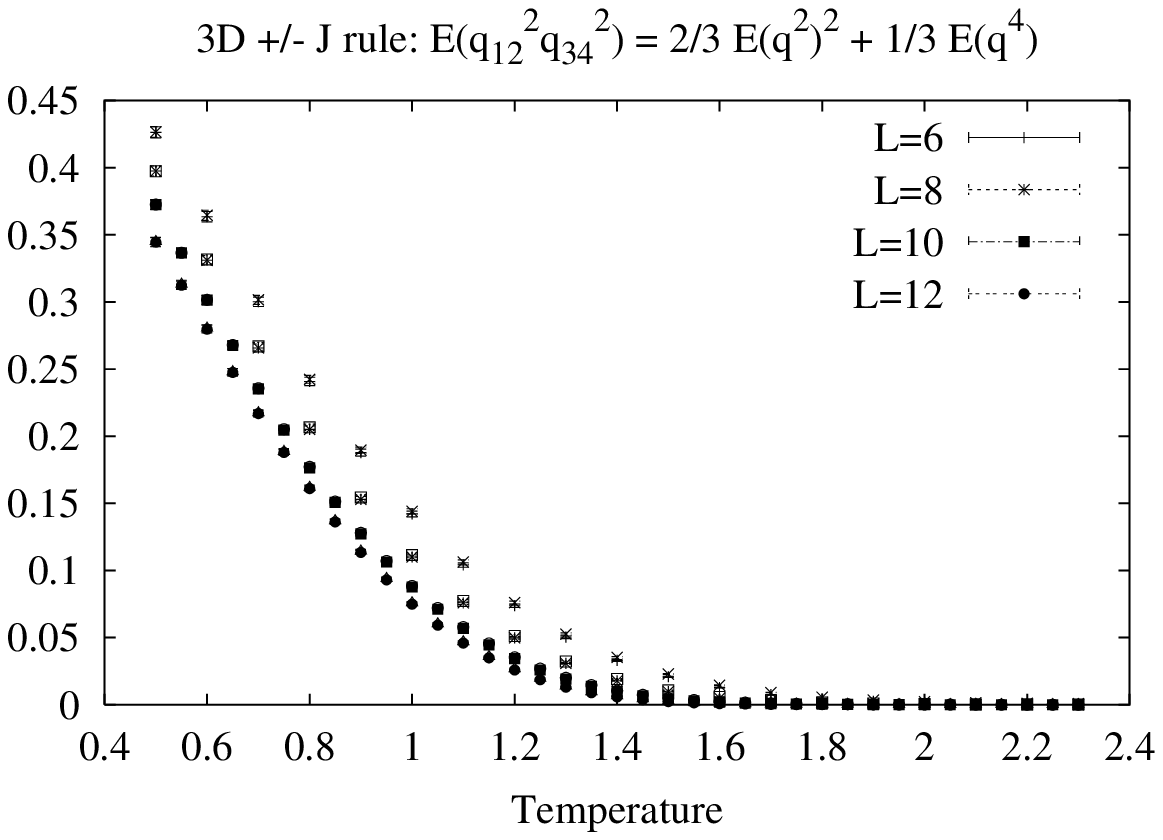}
\caption{The quantities $\ba{\lan q^2 \ran^{2}}$ {\bf and}  $\frac13 \ba{\lan q^4\ran}+
\frac23 \ba{\lan q^2\ran}^2$ as function of the temperature for different values of $L$ in $D=3$.}
\label{G2}
\end{figure}

We have tested the previous relations in three dimensions as function of the temperature 
at different values of $L$ \cite{CINQUE}.  In fig.  (\ref{G1}) we plot the quantity $\ba{\lan q^2 
\ran^{2}} - (\frac13 \ba{\lan q^4\ran}+ \frac23 \ba{\lan q^2\ran}^2)$, which should be equal 
to zero. Indeed it is very small and its values decreases with $L$.
In order to give a 
more precise  idea of the accuracy of stochastic stability in fig. (\ref{G2}) we plot 
separately the quantities $\ba{\lan q^2 \ran^{2}}$ {\bf and}  $\frac13 \ba{\lan q^4\ran}+
\frac23 \ba{\lan q^2\ran}^2$. The two quantities cannot be distinguished on this scale 
and in the low temperature region each of them is a factor of $10^{3}$ bigger of their 
difference. I believe that there should be few doubts on the fact that stochastic 
stability is satisfied for three dimensional spin glasses \cite{SEI} .

 \subsection{Overlap equivalence and ultrametricity}
 
In the case in which the overlap may take  three or more values, stochastic stability 
apparently does not fix the probability distribution. A further general principle may be 
useful in order to get new constraints. This principle is overlap equivalence.

In order to formulate the principle of overlap equivalence it is convenient to introduce a 
generalized overlap.  Let $A(i)$ be a local quantity. We define:

\be
q^{A}_{\al,\ga} =N^{-1}\sum_{i}\lan A(i) \ran_{\al}\lan A(i) \ran_{\ga} \ .
\ee
Let us consider two possibilities:
\begin{itemize}
    \item For $A(i)=\si(i)$ we get the usual overlap: $q^{A}= q$.
    \item For $A(i)=H(i)$ we get the usual energy overlap: $q^{A}= q^{E}$.
\end{itemize}

If we consider also the generalized overlaps we have that the description of a system is much more 
involved: we have to specify the $w_{\al}$ and the $q^{A}_{\al,\ga}$ for all possible choices of 
$A$.  In the general case an infinite number of quantities (i.e. $q^{A}_{\al,\ga}$, for all choices 
of $ A$) characterizes the mutual relations among the state $\al$ and the state $\ga$ 
\cite{FACE,CINQUE}.

Overlap equivalence states that this infinite number of different overlaps is reduced to one (the usual overlap) 
\cite{PARI,ULTRA,QUATTRO}.  There is only one significant overlap and all overlaps (depending on the 
operator $A$) are given functions of the spin overlap.  For any choices of $ A$ there is a 
corresponding function 
$f^{A}(q)$ such that $q^{A}_{\al,\ga}=f^{A}(q_{\al,\ga})$.

Overlap equivalence may be also formulated if we define the $q$-restricted ensemble:
\be
\lan f(\si,\tau) \ran_{q} \propto \sum_{\si.\tau}f(\si,\tau)  \delta(Nq-q[\si,\tau]) \ .
\ee
Overlap equivalence implies the validity of
cluster decomposition in the $q$-restricted ensembles.

Overlap equivalence (plus stochastic stability) seems \cite{ULTRA} to imply the ultrametricity condition 
\be
q_{\al,\ga} \ge \min(q_{\al,\beta},q_{\beta,\ga})\ \ \ \forall \beta.
\ee
If ultrametricity is valid, one finds that
\be
P^{12,23,31}(q_{12},q_{23},q_{31}) = 0 \quad ,
\ee
as soon as one of the three ultrametricity relations,
\be
q_{12} \ge \min(q_{23},q_{31}) \  ,  
q_{23} \ge \min(q_{31},q_{12}) \  ,  
q_{31} \ge \min(q_{12},q_{23}) \  , 
\ee
is not satisfied. 

It is remarkable that, given the function $P(q)$, the ultrametricity condition
completely determines the probability $P^{12,23,31}$, if we assume stochastic stability.
Overlap equivalence may be less compulsory of stochastic stability. There are some 
indications \cite{CAC} for its validity in the finite dimensional case (i.e. beyond mean field 
theory), but they are no so strong as 
for stochastic stability.

\section{The algebraic replica approach}

The algebraic replica approach is a compact way to code all the previous information into a 
matrix and also to compute the free energy \cite{MPV,PABOOK,PARI}.

A crucial role is played by a matrix $Q_{a,b}$ which is said to be a $0 \times 0$ matrix.  The 
direct definition of a $0 \times 0$ matrix may be not too easy.  Instead we can consider a family 
$Q_{a,b}^{(n)}$ of $n\times n$ matrices which depend {\sl analytically} on $n$: they are defined for 
$n$ multiple than $M$ in such a way that the analytic continuation of some scalar functions of these 
matrices at $n=0$ is well defined.

In this formulation the probability (after average over the permutation of 
lines and columns of the matrix) that an element of the matrix $Q_{a,b}$ with $a \ne b$ is equal to 
$q$, coincides with the function $P(q)\equiv \ba{P_{J}( q)}$:
\be
P(q)=\ba{ \sum_{\al,\ga}w_{\al}w_{\ga}\delta(q-q_{\al,\ga})}=
\lim_{n \to 0}{\sum_{a,b}\delta(Q_{a,b}-q)\over n(n-1)} \ .
\ee

In the same way the probability that an element of the matrix ($Q_{a,b}$) is equal to $q_{1}$ and 
an other element of the matrix ($Q_{c,d}$) is equal to $q_{2}$ (with $a$, $b$, $c$ and $d$ 
different) give us $P(q_{1},q_{2})\equiv \ba{P_{J}( q_{1}) P_{J}(q_{2})}$:
\be
P(q_{1},q_{2})=\ba{ \sum_{\al,\beta,\ga,\eps}w_{\al}w_{\beta}w_{\ga}w_{\eps}
\delta(q_{1}-q_{\al,\beta})\delta(q_{2}-q_{\ga,\eps})}=
\lim_{n \to 0}{\sum_{a,b,c,d}\delta(Q_{a,b}-q_{1})\delta(Q_{c,d}-q_{2})\over n(n-1)(n-2)(n-3) }\ .
\ee

In this approach probability statements become algebraic statements.
\begin{itemize}
\item
Stochastic stability becomes the statement that
one line of the matrix is a permutation of an other line of the matrix.
\item
Overlap equivalence is a  equivalent to more complex statement: if there are four indices ($a$. 
$b$, $c$ and $d$) such that $Q_{a,b}=Q_{c,d}$, there 
is a permutation $\pi$ that leaves the matrix unchanged ($Q_{a,b}=Q_{\pi(a),\pi(b)}$) and 
brings $a$ in $c$ and $b$ in $d$ (i.e. $\pi(a)=c$ and $\pi(b)=d$).
\end{itemize}

As you see in the algebraic approach one uses a quite different (and more abstract) language from the 
probabilistic approach. Using this language computations are often more simple and 
compact.

\section{Off-equilibrium dynamics}

The general problem that we face is to find what happens if the system is carried in a 
slightly off equilibrium situation. The are are two ways in which this can be done.
\begin{itemize}
    \item We rapidly cool the system starting from a random (high temperature) 
    configuration at time zero and we wait a time much larger than the microscopical one.  
    The system orders at distances smaller that a coherence distance $\xi(t)$ (which 
    eventually diverges when $t$ goes to infinity) but remains always disordered at 
    distances larger than $\xi(t)$.
    \item A second possibility consists in forcing the system in on off-equilibrium state  
    by gently {\it shaking} it. This can be done for example by adding a small time 
    dependent magnetic field, which should however strong enough to force a large scale 
    rearrangement of the system \cite{TRE}. 
\end{itemize}

In the first case we have the phenomenon of ageing. This effects may be evidenziated if we  
define  a two time 
correlation function and two time relaxation functions (we cool the system at time 0) 
\cite{cuku,frame}. 
The correlation function is defined to be
\be
C(t,t_w) \equiv \frac1N \sum_{i=1}^N \lan \si_i(t_w) \si_i(t_w+t)\ran\ ,
\ee
which is equal to the overlap $ q(t_{w},t_{w}+t)$ among a configuration at time $t_{w}$ and one at 
time $t_{w}+t$.
The relaxation function $S(t,t_w)$ is a just given by
\be
\beta ^{-1}\lim_{\delta h \to 0 } {\delta m(t+t_{w}) \over \delta h}\ ,
\ee
where $\delta m$ is the variation of the magnetization when we add  a magnetic 
field $\delta h$ starting from time $t_{w}$. More generally we can introduce the time dependent 
Hamiltonian:
\be
H=H_{0}+\theta(t-t_{w}) \sum_{i} h_{i}\si_{i} \ .
\ee
The relaxation function is thus defined as:
\be
\beta S(t,t_w) \equiv \frac1N \sum_{i=1}^N  \lan {\partial \si_i(t_w+t)\over \partial h_{i}} \ran\ .
\ee

\begin{figure}
    \centering
\includegraphics[width=0.7\textwidth]{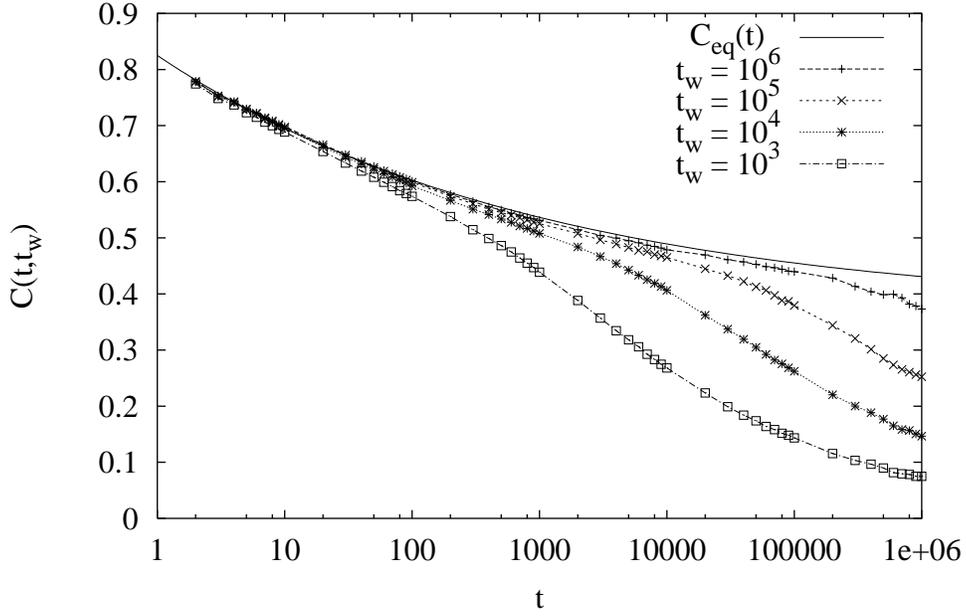}
\caption{The correlation function for spin glasses  as function of time $t$ at 
different $t_{w}$.}
\label{PLOT-CORR}
\end{figure}

We can distinguish two situations
\begin{itemize}
\item For $t<<t_w$ we stay in the quasi-equilibrium'' regime \cite{FV}, $C(t,t_w) \simeq C_{\rm 
eq}(t)$ , where $C_{\rm eq}(t)$ is the equilibrium correlation function; in this case $q_{EA} 
\equiv \lim_{t\to \infty} \lim_{t_w\to \infty} C(t,t_w)$.  

\item For $t=O(t_w)$ or larger we stay in 
the aging regime.  In the case where simple aging holds, $C(t,t_w) \propto {\cal C}(t/t_w)$.  A plot of the 
correlation function for spin glasses at different $t_{w}$ is shown in fig. (\ref{PLOT-CORR}).
\end{itemize}

\begin{figure}
    \centering
\includegraphics[width=0.5\textwidth]{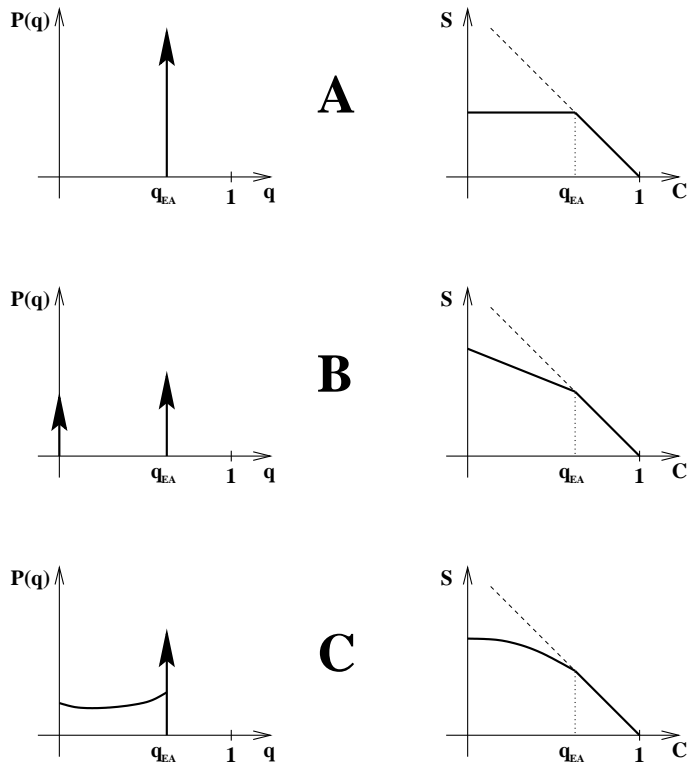}
\caption{Three different form of the function $P(q)$ and the related function $S(q)$. Delta 
functions are represented as a vertical arrow}
\label{CLASS}
\end{figure}

In the equilibrium regime, if we plot parametrically the relaxation function 
as function of the correlation, we find that
\be 
{ dS \over dC} =-1 \label{FDT}\ ,
\ee
which is a compact way of writing the fluctuation-dissipation theorem.

Generally speaking the fluctuation-dissipation theorem is not valid in the off-equilibrium regime.  
In this case one can use stochastic stability to derive a relation a among statics properties and 
the form of the function $S(C)$ measured in off-equilibrium \cite{cuku,frame,QUATTRO}:
\be
-{dS \over dC}=\int_{0}^{C} dq P(q)\equiv X(C) \ \label{FDR}.
\ee

In fig. (\ref{CLASS}) we show  three main different kinds of dynamical response $S(C)$, that correspond
to different shapes of the static $P(q)$ (which in the case of spin glasses at zero 
magnetic field should be replaced by $P(|q|)$). Case $A$ correspond to systems where replica symmetry 
is not broken, case B to one step replica symmetry breaking, which should be present in structural 
glasses and case $C$ to continuous replica symmetry breaking, which is present in spin glasses.

\begin{figure}
    \centering
\includegraphics[width=.6\textwidth]{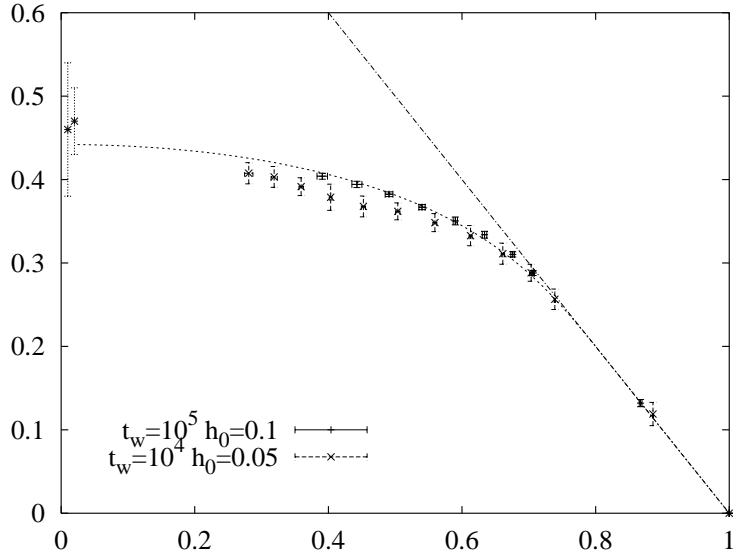}
\caption{Relaxation function versus correlation in the Edwards-Anderson (EA) model in $D=3$
     $T=0.7\simeq \frac34 T_c$ and theoretical predictions from eq. (\ref{FDR}).}
\label{EA3}
\end{figure}

The validity of these relation has been intensively checked in numerical experiments (see 
for example fig. (\ref{EA3})).

In spin glasses the relaxation function has been experimentally measured many times in the aging 
regime, while the correlation function has not yet been measured: it would be a much more difficult experiment 
in which one has to measure thermal fluctuations.  Fortunately enough measurements of both 
quantities for spin glasses are in now progress.  It would be extremely interesting to see if they 
agree with the theoretical predictions.

For reasons os time I shall not discuss the generalization of the previous arguments to the case of 
a spin glass in presence of a time dependent magnetic field.  I only remark that in this case the 
correlation function is directly related to the Birkhausen noise, which as far a I know, has never 
been measured in spin glasses.

\section{Conclusions}

In this talk I have presented a review of the basic ideas in the mean field approach to spin
glasses. There are many points which I have not covered and are very important.
Let me just mention some of them;
\begin{itemize}
    \item
    The analytic studies of the corrections to mean field theory.
    \item The purely dynamical approach which can be used without any reference to 
    equilibrium.
   \item
    The extension of these ideas to other disorder systems, to neural network and in 
    general to the problem of learning.
    \item
    The relevance of this approach for biological systems, both at the molecular level and 
    at the systemic level.
     \item
    The extension of these ideas to systems in which quenched disorder is absent, e.g. 
    structural
    glasses \cite{vetri-rsb}.
\end{itemize}

\section*{Acknowledgments}
I would like to especially thank those friend of mine with which I had a long standing collaboration 
though many years and among them to Silvio Franz, Enzo Marinari, Marc M\'ezard, Federico Ricci, Felix 
Ritort, Juan-Jesus Ruiz-Lorenzo, Nicolas Sourlas, Gerard Toulouse and Miguel Virasoro.  I would also 
like to express my debt of gratitude with all the people with have worked with me in the study of 
these problems and also to all others with whom I have not collaborated directly, but have been 
working in the same directions and have done important and significative progresses that have been 
crucial for reaching the present level of understanding.

\end{document}